\documentstyle[epsf,letters]{mn}  

\title[The colour--magnitude relation]{ The photoionization effect of
the ultraviolet background on the colour-magnitude relation of
elliptical galaxies}

\author[M. Nagashima and N. Gouda]{Masahiro Nagashima
\thanks{masa@th.nao.ac.jp} and Naoteru
Gouda\thanks{naoteru.gouda@nao.ac.jp}\\ National Astronomical
Observatory, Mitaka, Tokyo 181-8588, Japan\\ }

\begin{document}   
\maketitle   
\begin{abstract}   
  We examined effects of the ultraviolet background radiation (UVB) on
  the colour--magnitude relation (CMR) of elliptical galaxies in
  clusters of galaxies in the hierarchical clustering scenario by using
  a semi-analytic model of galaxy formation.  In our model the UVB
  photoionizes gas in dark haloes and suppresses the cooling of the
  diffuse hot gas onto galaxy discs.  By using a semi-analytic model
  without the effect of the UVB, Kauffmann \& Charlot found that the CMR
  can be reproduced by strong supernova heating because such supernova
  feedback suppresses the chemical enrichment in galaxies especially for
  small galaxies.  We find that the CMR also becomes bluer because of
  the UVB, in a different way from the effect of supernova feedback.
  While the supernova feedback suppresses the chemical enrichment by a
  similar mechanism to galactic wind, the UVB suppresses the cooling of
  the hot gas.  This fact induces the suppression of the metallicity of
  the intracluster medium (ICM).  In our model we find that the
  existence of the UVB can plausibly account for an observed ICM
  metallicity that is equal to nearly 0.3 times the solar value, and
  that in this case we can reproduce the CMR and the metallicity of the
  ICM simultaneously.
\end{abstract}   
   

\section{INTRODUCTION}   
It is well known that elliptical galaxies in clusters of galaxies have a
tight correlation between their colours and magnitudes (so-called
`colour--magnitude relation'; hereafter CMR).  For example, the rms
scatter about the mean CMR is typically $\sim$ 0.04 mag in the Virgo and
Coma clusters of galaxies.  It is equivalent to observational errors
(Bower, Lucey \& Ellis 1992).  Because it is believed that this tight
relation reflects formation and evolution processes of elliptical
galaxies, many people have studied this relationship in order to
understand galaxy formation processes.

The traditional scenario of formation of ellipticals is that a
monolithic protogalactic cloud collapses and then forms stars over a
short time-scale until the formation of a galactic wind (Larson 1974;
Arimoto \& Yoshii 1986, 1987).  In this framework, Kodama \& Arimoto
(1997) found that the CMR is corresponding to the sequence of the mean
stellar metallicities (the {\it metallicity sequence}) by comparing the
CMRs of high redshift cluster galaxies with those of theoretical models
based on the traditional scenario.  This fact that the CMR is a
metallicity sequence was also confirmed by investigating photometric
properties of the stellar populations of ellipticals (Ferreras, Charlot
\& Silk 1999).

In contrast to the collapse/wind model, recent developments on both
theory and observation of the cosmological structure formation are
revealing that objects such as galaxies and clusters of galaxies are
formed through hierarchical clustering of smaller objects.  Kauffmann \&
Charlot (1998; hereafter KC) applied their semi-analytic model to the
problem of the CMR and reproduced the observed CMR when the model
includes the chemical evolution process and strong feedback to
interstellar media by supernovae.  Moreover, they found a tight
luminosity--metallicity relation and reproduced other properties of
cluster ellipticals such as the line indices and the Faber-Jackson
relation.

KC interpreted why the CMR is reproduced when the feedback process is
strong as follows; they considered that giant ellipticals must be formed
by mergers of larger spiral galaxies and that dwarf ellipticals must be
formed from smaller spirals.  Because the feedback strength depends on
the mass of galaxies, the metallicity of stars in more massive
progenitors becomes higher than that in smaller progenitors.  Therefore
giant ellipticals become redder than dwarf ellipticals.

In this work, we introduce the photoionization process of galactic gas
by the UV background radiation (UVB) in our model.  We have already
investigated the effect of the UVB on the luminosity function and the
colour distribution of galaxies (Nagashima, Gouda \& Sugiura 1999), and
here investigate this effect on the CMR, which has not been considered.
Moreover, it is also shown how the metal abundance in the intracluster
medium (ICM) is affected by the UVB.  Through this work, the difference
between the mechanisms and effects of the supernova feedback and the UVB
in the galaxy formation process will be clarified.

In Section \ref{sec:model}, we describe the semi-analytic model used
here briefly.  In Section \ref{sec:cmr}, we show the CMR and show that
the effect of the UVB seems similar to the effect of the supernova
feedback.  In Section \ref{sec:chem}, we investigate the property of the
effect of the UVB on the CMR.  We will show that the CMR hardly depends
on the intensity of the UVB if it is sufficiently strong, and interpret
this result physically by using a simple model.  In Section 5, the metal
abundance of the ICM is discussed.  Section \ref{sec:conc} is devoted to
conclusions.

\section{MODEL}\label{sec:model}
Here we describe the semi-analytic model that we use briefly.  At first,
merging histories of dark haloes are realized by the extension of the
Press-Schechter formalism (Press \& Schechter 1974).  We adopt a method
given by Somerville \& Kolatt (1999).  The cosmological model is fixed
to a cosmological constant-dominated flat universe, $\Omega_{0}=0.3,
\Omega_{\Lambda}=0.7, h=0.7$ and $\sigma_{8}=1$, where $h$ is the Hubble
parameter, $h\equiv H_{0}/100$km~s$^{-1}$Mpc$^{-1}$, and $\sigma_{8}$ is
the normalization of the power spectrum of density fluctuation.  We use
a power spectrum given by Bardeen et al. (1986).  Haloes with circular
velocity smaller than 40km~s$^{-1}$ are treated as diffuse matter.  In
this paper, we consider haloes only with circular velocity $V_{\rm
c}=10^{3}$ km~s$^{-1}$ at $z=0$, which is a typical value for a cluster
of galaxies.

Next, in the merging path, we calculate the evolution of the baryonic
component from higher redshift to present.  Diffuse gas in a newly
collapsing dark halo is heated up to the virial temperature of the dark
halo by shock heating (the {\it hot gas}).  The shock heating occurs
only in the case that the mass ratio of the largest progenitor among all
progenitors of the halo to the newly collapsing halo exceeds $f_{\rm
reheat}$.  In the other case, the hot gas in the new halo conserves the
temperature of the largest progenitor.  If the new halo has no
progenitor halo, the mass fraction of the hot diffuse gas is equal to
$\Omega_{\rm b}/\Omega_{0}$.  Here we use $\Omega_{\rm b}=0.015h^{-2}$.
Assuming the singular isothermal density distribution of the hot gas, we
calculate the cooled gas mass by equating the cooling time-scale
$\tau_{\rm cool}(r)$ to the elapsed time from the last shock heating
$t_{\rm elapse}$, $\tau_{\rm cool}(r_{\rm cool})=t_{\rm elapse}$, where
the cooling function given by Sutherland \& Dopita (1993) is used.  If
the UVB does not exist, the gas within the cooling radius, $r_{\rm
cool}$, cools and is accreted by a disc of the halo central galaxy.  In
the other case, the UV photons penetrate the cooled gas from $r_{\rm
cool}$ to $r_{\rm UV}$, where $r_{\rm cool} > r_{\rm UV}$, and the outer
layer $r>r_{\rm UV}$ is photoionized.  This $r_{\rm UV}$ is estimated by
assuming the inverse Str{\"o}mgren sphere approximation (Nagashima et
al. 1999).  Thus the gas within $r_{\rm UV}$ is cooled.  In this paper,
we assume that the intensity of the UVB, $J$, evolves as $J\propto
(1+z)^{\gamma}$, where $\gamma=4$ for $z\leq 2$ and $\gamma=-1$ for
$2<z\leq 6$.  The UVB does not exist before $z=6$.  We found that the
results in this work does not depend on the shape of the intensity
evolution of the UVB qualitatively.  In the following, we parameterize
the UV intensity at $z=2$ by $J_{-21}\equiv
(J/10^{-21}\mbox{erg~cm}^{-2}\mbox{s}^{-1}\mbox{Hz}^{-1}\mbox{sr}^{-1})$.
It should be noted that in the case where the circular velocity of
haloes exceeds 500 km s$^{-1}$, we prevent the cooling process manually.
This is the same procedure as Kauffmann, White \& Guiderdoni (1993).

Stars are formed from this cold gas.  The star formation rate in discs
is given by $\dot{M}_{*}=M_{\rm cold}/\tau_{*}$, where $M_{*}$ and
$M_{\rm cold}$ are the masses of the stars and cold gas, respectively,
and $\tau_{*}$ is the star formation time-scale.  We adopt a similar
form to that given by KC, $\tau_{*}=\tau_{*}^{0}[\tau_{\rm
dyn}/\tau_{\rm dyn}(z=0)]$, where $\tau_{\rm dyn}$ is the dynamical
time-scale of the halo which is calculated by the spherical collapse
approximation (Tomita 1969; Gunn \& Gott 1972), and $\tau_{*}^{0}$ is a
free parameter, which is set to 2Gyr in this paper.

With star formation, supernovae occur and heat up the surrounding cold
gas to the hot phase (supernova feedback).  The reheating rate is given
by $\dot{M}_{\rm reheat}=\beta(V_{\rm c})\dot{M}_{*}$, where
$\beta(V_{\rm c})=(V_{\rm c}/V_{\rm hot})^{-\alpha_{\rm hot}}$, $V_{\rm
hot}$ and $\alpha_{\rm hot}$ are free parameters, respectively.  This is
the same parameterization given by Cole et al. (1994), and we adopt
$\alpha_{\rm hot}=2$ according to KC.

Chemical enrichment is solved in a consistent way to the above star
formation process.  The yield $y$ is assumed to be twice the solar value
according to the strong feedback model of KC.  While KC assumed that a
fraction $f$ of metals is released from ejecta of the supernovae to the
hot gas directly, we do not assume such a process.  The metals are
released to the hot gas by the supernova feedback in the same way as for
the cold gas.

We recognize a system consisting of the stars and cooled gas as a {\it
galaxy}.  Hereafter we define the mass of a galaxy as the mass of stars
and cold gas of the galaxy.  When two or more dark haloes merge
together, there is a possibility that galaxies contained in progenitor
haloes merge together.  We define a central galaxy as the central galaxy
of the largest progenitor halo.  Other galaxies are defined as satellite
galaxies.  When the elapsed time of a galaxy from the epoch at which the
galaxy becomes satellite exceeds the dynamical friction time-scale of
the halo, the galaxy merges with the central galaxy of the halo.  When
galaxies merge together, if the mass ratio of the smaller galaxy to the
larger galaxy exceeds $f_{\rm bulge}$, starburst occurs and all cold gas
is consumed.  The same feedback law as that for discs is adopted.  Then,
all of stars becomes bulge stars.  In the other case, the smaller galaxy
merges into the disc component of the larger galaxy without any
additional star formation activity.  In this paper we adopt $f_{\rm
bulge}=0.2$.

Finally, we calculate the colour and luminosity of each galaxy from star
formation history of each galaxy by using the stellar population
synthesis technique.  We use a simple stellar population model given by
Kodama \& Arimoto (1997).  Through the above procedures, we obtain a
colour-magnitude diagram of galaxies.  In the follows, we pick out
galaxies with the $B$-band bulge-to-disc luminosity ratio larger than
1.52 as elliptical galaxies according to Simien \& de Vaucouleurs
(1986).

The above procedure is mainly based on Kauffmann et al. (1993), Cole et
al. (1994, 2000) and Somerville \& Primack (1999).  Though the details
of our model are slightly different from the previous ones, we confirmed
that the main results of the previous work, such as both of the field
and cluster luminosity functions, colour distribution and cold gas mass
fraction of galaxies, are almost reproduced by our model within similar
range of values of the parameters (Nagashima \& Gouda, in preparation).

\section{COLOUR-MAGNITUDE RELATION}\label{sec:cmr}
\begin{figure}
\epsfxsize=\hsize
\epsfbox{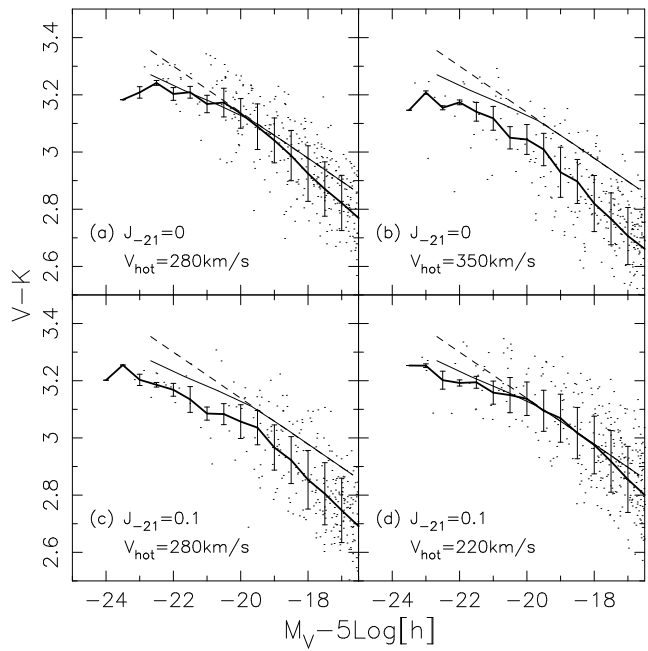}

\caption{$V-K$ Colour--magnitude diagram.  The thick solid curves with
 errorbars denote the model CMRs averaged over 50 clusters and the
 errorbars denote the mean values of the 1$\sigma$ scatter for each
 realization.  Each dot indicates an elliptical galaxy only for five
 realizations.  The thin dashed line in each panel indicates the CMR in
 the Coma cluster (Bower et al. 1992).  The thin solid line denotes the
 aperture-corrected CMR in the Coma (Kodama et al. 1998).  Model
 parameters for the four models are shown in Table \ref{tab:icm}.  Upper
 panels: $J_{-21}=0$.  (a) $V_{\rm hot}=280$ km~s$^{-1}$.  (b) $V_{\rm
 hot}=350$ km~s$^{-1}$.  Lower panels: $J_{-21}=0.1$.  (c) $V_{\rm
 hot}=280$ km~s$^{-1}$.  (d) $V_{\rm hot}=220$ km~s$^{-1}$.  Note that
 the simulations presented in (a) and (c) have the same strength of the
 supernova feedback.}

\label{fig:fig1}
\end{figure}

In Figure \ref{fig:fig1}, we show the colour-magnitude diagram for four
models.  The thick solid curves with errorbars indicate the CMRs given
by the models averaged over 50 realizations.  The errorbars denote the
mean values of 1$\sigma$ scatter for each realization, so they are
corresponding to the scatter for each cluster of galaxies.  Dots
indicate ellipticals only for five realizations.  The thin dashed lines
indicate the observational CMR in the Coma cluster (Bower, Lucey \&
Ellis 1992) and the thin solid lines denote the aperture-corrected CMR
in the Coma cluster (Kodama, Arimoto, Barger \& Arag{\'o}n-Salamanca
1998).  Since colour of each elliptical galaxy in our model is
integrated over the whole region of the galaxy, the model CMR should be
compared with the aperture-corrected CMR.  Upper panels show the models
without the UVB and lower panels show those with the UVB of
$J_{-21}=0.1$.

As shown by KC, when we do not consider the effect of the UVB, the model
with $V_{\rm hot}=280$ km~s$^{-1}$ can reproduce the observed CMR
(Figure \ref{fig:fig1}a).  Imposing stronger feedback with $V_{\rm
hot}=350$ km~s$^{-1}$, the colours of ellipticals become too blue to be
consistent with the observation (Figure \ref{fig:fig1}b) because the
metallicity of ellipticals decreases (see next section).  The same
effect of the bluing is also caused by the UVB.  In Figure
\ref{fig:fig1}c, we show the model with the same value of $V_{\rm hot}$
as that of the model in Figure \ref{fig:fig1}a and with the UVB of
$J_{-21}=0.1$.  By weakening the feedback strength from $V_{\rm
hot}=280$ to 220km~s$^{-1}$, we obtain a CMR in agreement with the
observation again (Figure \ref{fig:fig1}d).  This manipulation is
corresponding to reverse of the procedure from Figure \ref{fig:fig1}a to
\ref{fig:fig1}b.  Thus we obtain a similar CMR to the observation with
both of the weaker feedback and the UVB.

It should be noted that KC found that the CMR reflects the
metallicity-luminosity relation.  We also find that models with similar
CMR show similar metallicity-luminosity relation independent of the
existence of the UVB, e.g., the differences of the
metallicity-luminosity relations between the models of Figures 1a and 1d
and between Figures 1b and 1c are negligible.  Thus the above properties
of the CMR, i.e., the dependences on the supernova feedback and on the
UVB, are related with the stellar metallicity of ellipticals.
In this connection, we find that the shape of the metallicity-luminosity
relation of the four models is similar to the metallicity sequence of
Kodama \& Arimoto (1997), in spite of the significant difference between
our model and their monolithic cloud collapse model.  This also suggests
the importance of investigating the chemical enrichment process of
galaxies.

\section{CHEMICAL ENRICHMENT OF GALAXIES}\label{sec:chem}
\begin{figure}
\epsfxsize=\hsize
\epsfbox{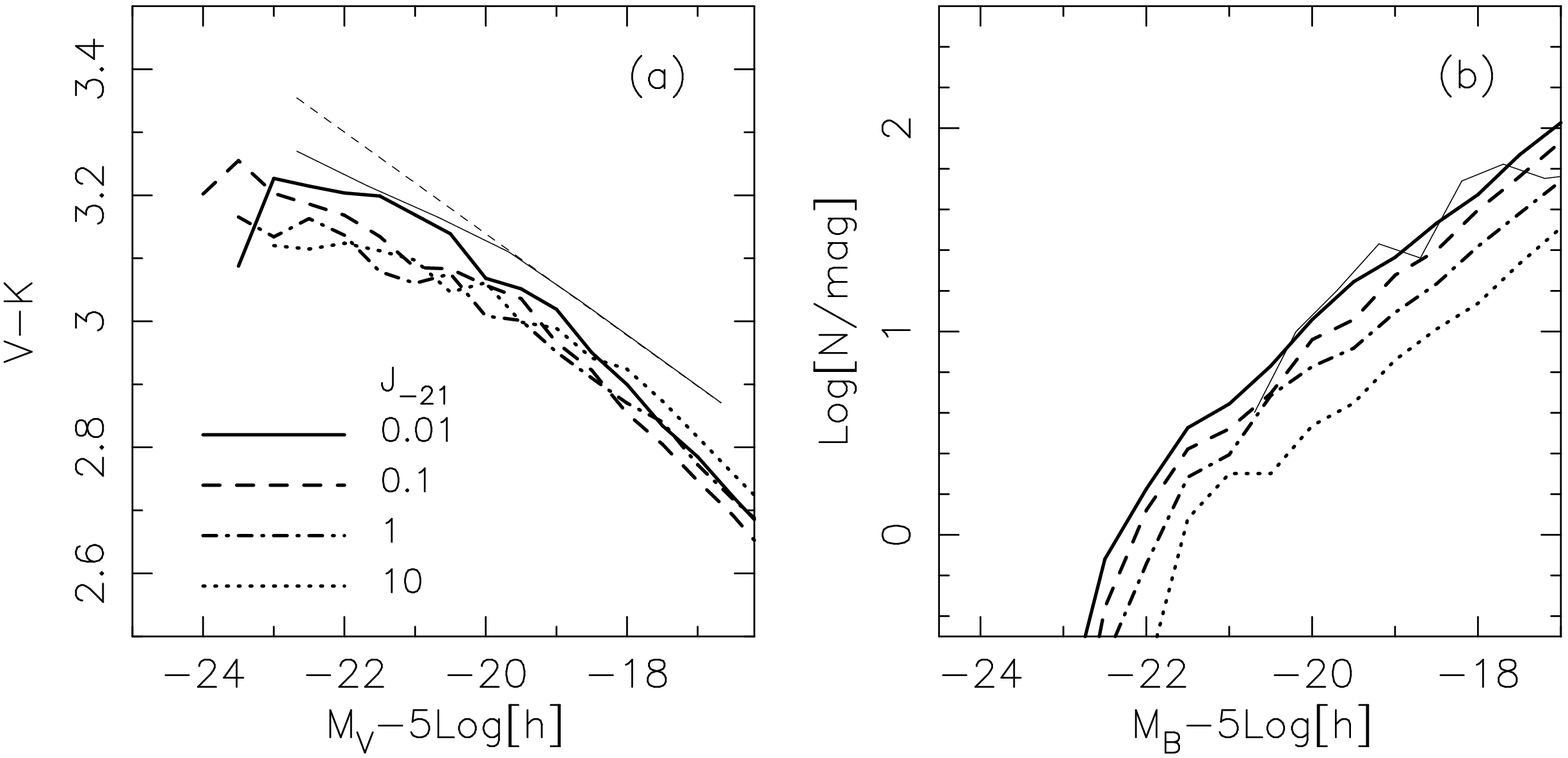}

\caption{(a) Colour--magnitude relations.  Model parameters of these
 four models are the same as those of Figure \ref{fig:fig1}c ($V_{\rm
 hot}=280$ km~s$^{-1}$) but for the intensity of the UVB.  The solid,
 dashed, dot-dashed and dotted lines denote the models with
 $J_{-21}=0.01, 0.1, 1$ and 10, respectively.  (b) Cluster luminosity
 functions.  The types of lines are the same as (a).  The thin line is
 the luminosity function of the Virgo cluster galaxies by Sandage et
 al. (1985).}

\label{fig:fig2}
\end{figure}

While the above result seems to show that the effect of the UVB is the
same as the supernova feedback apparently, the ways to make ellipticals
metal-poor are different.  In Figure \ref{fig:fig2}a, we show four CMRs
with different strengths of the UVB, $J_{-21}=0.01, 0.1, 1$ and $10$ and
with the same feedback strength, $V_{\rm hot}=280$ km~s$^{-1}$.  The
CMRs are hardly changed by the UVB if $J_{-21}\geq 0.1$.  On the other
hand, the luminosity functions are clearly affected by the UVB.  In
Figure \ref{fig:fig2}b, we show the cluster luminosity functions of
these models.  The thin solid line denotes the observational cluster
luminosity function in the Virgo cluster given by Sandage et al. (1985).
These figures show that the sufficiently strong UVB decreases the number
of galaxies by its photoionization effect while it does not affect the
metallicity of each elliptical galaxy.

In the following, we explore the reason why the metallicity does not
depend on $J_{-21}$.  First of all, we evaluate the effect of the
supernova feedback.  The mass-weighted mean metallicity of each galaxy
is given by (Nagashima \& Gouda, in preparation)
\begin{eqnarray}
\langle
\lefteqn{Z_{*}(t)\rangle_{M_{*}}=}\nonumber\\
&&1-F
\frac{1-\exp\left[-(1+\beta-(R-\alpha
y)(1-f))\frac{t-t_{s}}{\tau_{*}}\right]}
{1-\exp\left[-(1+\beta-R(1-f))\frac{t-t_{s}}{\tau_{*}}\right]},
\label{eqn:meanstz}
\end{eqnarray}
where
\begin{equation}
F=\frac{1-Z_{\rm cool}^{0}}{1+\frac{(1-f)\alpha y}{1+\beta-R(1-f)}},
\label{eqn:meanstz2}
\end{equation}
$\alpha$ is a locked-up mass fraction, $\alpha=1-R$ ($R$ is the gas
fraction returned by evolved stars), and $Z_{\rm cool}^{0}$ is an
initial metallicity of cold gas at $t=t_{s}$, where $t_{\rm s}$ is an
initial time.  In this paper, $f=0$ and $R=0.25$.  The starburst
corresponds to $t/\tau_{*}\to\infty$, so the final stellar metallicity
depends only on the strength of the supernova feedback, $\beta$, as
\begin{eqnarray}
\langle Z_{*}(t\to\infty)\rangle_{M_{*}}&=&\frac{(1-Z_{\rm cool}^{0})\alpha y}{1+\beta-R+\alpha y}+Z_{\rm cool}^{0}\nonumber\\
&\simeq&\frac{y}{1+\beta}+Z_{\rm cool}^{0}.
\end{eqnarray}
Therefore the stronger the feedback is, the lower the metallicity is.
Thus the differences between Figures \ref{fig:fig1}a and \ref{fig:fig1}b
and between Figures \ref{fig:fig1}c and \ref{fig:fig1}d are explained by
the strength of the supernova feedback, $\beta$.

Next we investigate the process by which the UVB suppresses the
metallicity.  When the UVB exists, the amount of the cooled gas
decreases.  Then the amount of the metal polluted reheated gas also
decreases and the metallicity of the hot gas is suppressed.  The hot
gas, including metal polluted reheated gas, cools at the later stage
again.  Therefore the metallicity of the re-cooled gas decreases.  This
corresponds to decreasing $Z_{\rm cool}^{0}$ at the later stage in
eq.(3).  Thus the mean stellar metallicity is suppressed by the UVB.  If
the UVB is sufficiently strong ($J_{-21}\ga 0.1$), then $Z_{\rm
cool}^{0}$ becomes negligible compared with the first term in eq.(3),
$y/(1+\beta)$.  As a result, the metallicity does not depend on the
$J_{-21}$.

\section{METAL ABUNDANCE OF INTRACLUSTER MEDIUM}
\begin{figure}
\epsfxsize=\hsize
\epsfbox{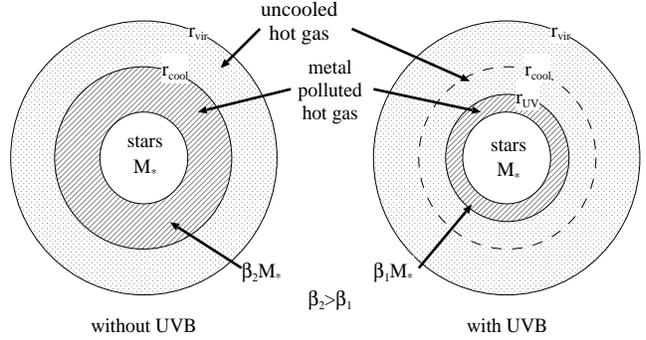}

\caption{Schematic descriptions of ellipticals in dark haloes.  These
 two galaxies have the same masses of dark matter and stars.  $r_{\rm
 vir}$ denotes the virial radius of the haloes.  The mass of the
 reheated, metal-polluted hot gas is $\beta M_{*}$.  The right model has
 weaker supernova feedback than the left model because of the UVB.  }

\label{fig:fig3}
\end{figure}

As considered in the previous section, the UVB may affect the
metallicity of the hot gas.  Here we investigate it by a simple model.
Now consider two ellipticals with the same stellar mass which are
illustrated in Figure \ref{fig:fig3}.  One is exposed to the UVB (right
in Figure \ref{fig:fig3}) and the other is not (left).  The second case
is required to have stronger feedback ($\beta_{2}$) than the first
($\beta_{1}$) because the same mass of stars must be formed from larger
amount of cold gas than in the first case.  Thus the second elliptical
has larger amount of the metal polluted reheated gas than the first one.

From the above discussion, we expect that the metal abundance of the ICM
of the first elliptical should become smaller than that of the second
elliptical.  We show the metallicity of the ICM, $Z_{\rm ICM}$, in Table
\ref{tab:icm} for the four models in Figure \ref{fig:fig1}.  $Z_{\rm
ICM}$ decreases significantly when the UVB exists.  Because of $Z_{\rm
ICM}\simeq 0.3Z_{\odot}$ in some observations (e.g., Tamura et
al. 2001), the existence of the UVB can plausibly account for the
metallicity of the ICM.  It should be noted that KC also obtained
$Z_{\rm ICM}\simeq 0.3$ with a non-zero value of $f$ without the effect
of the UVB.  Their model is nearly corresponding to our model with
$f_{\rm reheat}\to 0$, in which we find $Z_{\rm ICM}\simeq 0.3$ with no
UVB.  Thus our model is not inconsistent with KC.  At present there is
no method of determining the value of the parameter $f_{\rm reheat}$,
although Salvador-Sol{\'e}, Solanes \& Manrique (1998) suggest $f_{\rm
reheat}\sim 0.6$ by comparing the so-called universal density profile of
dark haloes (Navarro, Frenk \& White 1996) with the theoretical
prediction using both of the spherical collapse model and the formation
redshift distribution of haloes.  Thus we consider that models with
$f_{\rm reheat}\simeq 0.5$ are more realistic.

\begin{table}
\begin{center}  
\caption{Metallicity of intracluster medium}
\label{tab:icm}
\begin{tabular}{cccc}
\hline
Model & $V_{\rm hot}$ (km~s$^{-1}$) & $J_{-21}$ & $Z_{\rm ICM}/Z_{\odot}$\\
\hline
Figure 1a & 280 & 0   & 0.59$\pm$ 0.01 \\
Figure 1b & 350 & 0   & 0.49$\pm$ 0.01\\
Figure 1c & 280 & 0.1 & 0.34$\pm$ 0.02\\
Figure 1d & 220 & 0.1 & 0.38$\pm$ 0.02\\
\hline
\end{tabular}
\end{center}
\end{table}

\section{CONCLUSIONS}\label{sec:conc}
We investigate the effect of the UVB on the CMR in the hierarchical
clustering scenario by using a semi-analytic galaxy formation model, in
which we introduce the photoionization effect of the gas by the UVB.  We
show that the UVB plays a similar role on the CMR to the supernova
feedback apparently.  Both effects suppress the mean stellar metallicity
of each elliptical galaxy.  Therefore the CMR becomes bluer in the both
cases.

We find that the CMR is affected even if the UV intensity is weak,
$J_{-21}\sim 0.1$ and hardly depends on the intensity in the case of
sufficiently strong UVB, $J_{-21}\ga 0.1$, in contrast to the supernova
feedback.  This is because the UVB suppresses only the cooling process
(see below).  Moreover we show that the two mechanisms which degenerate
in the CMR can be distinguished by investigating the metallicity of the
ICM, and that the existence of the UVB is favoured to account for the
observed metal abundance of the ICM.  It will be possible to detect the
difference in the metallicity by observational studies with {\it
XMM-Newton}.

The physical mechanisms affecting the CMR of the UVB and of the
supernova feedback are different, as follows.  In the case of the strong
feedback without the UVB, while much cold gas is reheated before
metal-rich stars are formed, much metal-polluted gas is expelled in the
form of hot gas such as the galactic wind.  On the other hand, the UVB
suppresses gas cooling.  So the material forming stars, that is, the
cooled gas, is less than in the former no UVB case.  Under the UVB
scenario, if the supernova feedback is weaker than the former no UVB
case, a similar mass of stars can be formed from a smaller amount of the
cooled gas, and then the stars are chemically enriched.  Note that in
the case of a UVB the hot gas becomes metal-poor because a smaller
fraction of metals is expelled from galaxies than in the former case.
Thus conserving the stellar metallicity of each galaxy by introducing
the UVB and weakening the feedback, the metal abundance in the ICM can
become lower compared to the former case of no UVB.

Recently an overcooling problem has been discussed by using
high-resolution hydrodynamical simulations (Balogh et al. 2001), in
which too much gas cools compared with observations.  The effect of the
UVB, as well as strong supernova feedback, will be useful to help solve
this problem.  Besides by using the proximity effect, it is suggested
$J_{-21}\ga 1$ (Jennifer et al. 2000). In this paper, we showed that the
UVB affects the CMR and the ICM even in the case of $J_{-21}\ga 0.1$,
but considered the effect of the UVB only by introducing the simple
inverse Str{\"o}mgren sphere approximation.  Hence these results mean
that we should investigate the effect of the UVB in more detail.  While
we believe that we can understand the effect of the UVB qualitatively,
detailed radiative transfer calculation will be required for
quantitative estimation of the effect.

\section*{ACKNOWLEDGMENTS}    
We wish to thank K. Okoshi, F. Takahara, Y. Fujita, T. Kodama, and
N. Arimoto for useful suggestions.  We also thank the anonymous referee
who led us to a substantial improvement of our paper.  This work has
been supported in part by the Grant-in-Aid for the Scientific Research
Fund (10640229) of the Ministry of Education, Science, Sports and
Culture of Japan.

\bsp

\begin{thebibliography}{}   
\bibitem{}Arimoto N., Yoshii Y., 1986, A\&A, 164, 260
\bibitem{}Arimoto N., Yoshii Y., 1987, A\&A, 173, 23
\bibitem{}Balogh M.L., Pearce F.R., Bower R.G., Kay S.T., preprint,
	astro-ph/0104041
\bibitem{}Bardeen J.M., Bond J.R., Kaiser N., Szalay A.S., 1986, ApJ,
304, 15
\bibitem{}Bower R., Lucey J.R., Ellis R.S., 1992, MNRAS, 254, 601
\bibitem{}Cole S., Aragon-Salamanca A., Frenk C.S., Navarro J.F., Zepf
S.E., 1994, MNRAS, 271, 781
\bibitem{}Cole S., Lacey C.G., Baugh C.M., Frenk C.S., 2000, MNRAS, 319, 168
\bibitem{}Ferreras I., Charlot S., Silk J., 1999, ApJ, 521, 81
\bibitem{}Gunn J.E., Gott J.R., 1972, ApJ, 176, 1
\bibitem{}Jennifer S., Jill B., Adam D., Kulkarni V.P., 2000, ApJS, 130, 67
\bibitem{}Kauffmann G., White S.D.M., Guiderdoni B., 1993, MNRAS, 264, 201
\bibitem{}Kauffmann G., Charlot S., 1998, MNRAS, 294, 705 (KC)
\bibitem{}Kodama T., Arimoto N., 1997, A\&A, 320, 41
\bibitem{}Kodama T., Arimoto N., Barger A.J., Arag{\'o}n-Salamanca A.,
	1998, A\&A, 334, 99
\bibitem{}Larson R.B., 1974, MNRAS, 166, 585
\bibitem{}Nagashima M., Gouda N., Sugiura N., 1999, MNRAS, 305, 449
\bibitem{}Navarro J.F., Frenk C.S., White S.D.M., 1996, ApJ, 462, 563
\bibitem{}Press W.H., Schechter P., 1974, ApJ, 187, 425
\bibitem{}Salvador-Sol{\'e} E., Solanes J.M., Manrique A., 1998, ApJ,
	499, 542
\bibitem{}Sandage A., Binggeli B., Tammann G.A., 1985, AJ, 90, 1759
\bibitem{}Simien F., de Vaucouleurs G., 1986, ApJ, 302, 564
\bibitem{}Somerville R.S., Kolatt T., 1999, MNRAS, 305, 1
\bibitem{}Somerville R.S., Primack J.R., 1999, MNRAS, 310, 1087
\bibitem{}Sutherland R., Dopita M.A., 1993, ApJS, 88, 253 
\bibitem{}Tamura T. et al., 2001, A\&A, 365, L87
\bibitem{}Tomita K., 1969, PTP, 42, 9
\end{thebibliography}
\end{document}